\documentclass[10pt]{IEEEtran}
\usepackage{times,epsfig}
\usepackage{mathrsfs}
\usepackage{cite}
\usepackage{epsfig}
\usepackage{graphicx}
\usepackage{url}
\usepackage{stfloats}
\usepackage{amsmath}
\usepackage{array}
\usepackage{amssymb}
\usepackage{amsthm}
\usepackage{multirow}
\usepackage{cite}
\usepackage{algorithm}
\usepackage{algorithmic}

\newcommand{\minitab}[2][l]{\begin{tabular}{#1}#2\end{tabular}}

\ifCLASSINFOpdf
\else
\fi

\begin{document}

\bstctlcite{IEEEexample:BSTcontrol}

\title{Audio Splicing Detection and Localization Using Environmental Signature}

\author{Hong Zhao, Yifan Chen \IEEEmembership{Member,~IEEE}, Rui Wang \IEEEmembership{Member,~IEEE}, and Hafiz~Malik \IEEEmembership{Member,~IEEE,}

\thanks{Copyright (c) 2014 IEEE. Personal use of this material is permitted. However, permission to use this material for any other purposes must be obtained from the IEEE by sending a request to pubs-permissions@ieee.org}

\thanks{Hong Zhao, Yifan Chen and Rui Wang are with the Department of Electrical and Electronic Engineering, South University of Science and Technology of P.R. China. email: zhao.h@sustc.edu.cn, chen.yf@sustc.edu.cn, wang.r@sustc.edu.cn}

\thanks{Hafiz Malik is with the Department of Electrical and Computer Engineering, The University of Michigan -- Dearborn, MI 48128; ph: +1-313-593-5677; email: hafiz@umich.edu}}

\maketitle

\begin{abstract}
%\boldmath
Audio splicing is one of the most common manipulation techniques in the area of audio forensics. In this paper, the magnitudes of acoustic channel impulse response and ambient noise are proposed as the environmental signature. Specifically, the spliced audio segments are detected according to the magnitude correlation between the query frames and reference frames via a statically optimal threshold. The detection accuracy is further refined by comparing the adjacent frames. The effectiveness of the proposed method is tested on two data sets. One is generated from TIMIT database, and the other one is made in four acoustic environments using a commercial grade microphones. Experimental results show that the proposed method not only detects the presence of spliced frames, but also localizes the forgery segments with near perfect accuracy. Comparison results illustrate that the identification accuracy of the proposed scheme is higher than the previous schemes. In addition, experimental results also show that the proposed scheme is robust to MP3 compression attack, which is also superior to the previous works.
\end{abstract}


% \fi
%\IEEEpeerreviewmaketitle

\section{Introduction}
\IEEEPARstart{D}{igial} media (audio, video, and image) has become a dominant evidence in litigation and criminal justice, as it can be easily obtained by smart phones and carry-on cameras. Before digital media could be admitted as evidence in a court of law, its authenticity and integrity must be verified. However, the availability of powerful, sophisticated, low-cost and easy-to-use digital media manipulation tools has rendered the integrity authentication of digital media a challenging issue. For instance, \emph{audio splicing} is one of the most popular and easy-to-do attack, where target audio is assembled by splicing segments from multiple audio recordings. Thus, in this context,
we are trying to provide authoritative answers to the following questions, which should be addressed such as\cite{Hong:2013a}:
\begin{itemize}
  \item Was the evidentiary recording captured at location, as claimed?
  \item Is an evidentiary recording ``\emph{original}'' or was it created by splicing multiple recordings together?
\end{itemize}

In our previous works \cite{malik:2010,malik:2012a,zhao:2012,Hong:2013a}, we have shown that acoustic reverberation and ambient noise can be used for acoustic environment identification. One of the limitations of these methods is that these methods cannot be used for splicing location identification. And also, researchers have proposed various methods for the applications of audio source identification\cite{Hajj-Ahmad:2013}, acquisition device identification\cite{Panagakis:2013}, double compression\cite{Korycki:2013} and etc. However, applicability of these method for audio splicing detection is missing.

This paper presents a novel approach to provide evidence of the location information of the captured audio. The magnitudes of acoustic channel impulse response and ambient noise are used as intrinsic acoustic environment signature for splicing detection and splicing location identification. The proposed scheme firstly extracts the magnitudes of channel impulse response and ambient noise by applying the spectrum classification technique to each frame. Then, the correlation between the magnitudes of the query frame and the reference frame is calculated. The spliced frames are detected by comparing the correlation coefficients to the pre-determined threshold. A refining step using the relationship between adjacent frames is adopted to reduce the detection errors. One of the advantages of the proposed approach is that it does not depend on any assumptions as in \cite{Hong:2013a}. In addition, the proposed method is robust to lossy compression attack.

Compared with the existing literatures, the contribution of this paper includes: 1) Environment dependent signatures are exploited for audio splicing detection; 2) Similarity between adjacent frames is proposed to reduce the detection and localization errors.

%Here we exploit specific artifacts introduced at the time of recording as an intrinsic signature and for audio recording integrity authentication. Both the acoustic channel impulse response and the ambient noise are jointly considered to achieve this objective. In this scheme, each test audio is divided into overlapping frames. For each frame, the magnitudes of channel impulse response and ambient noise are jointly estimated as an intrinsic signature by applying the spectrum classification technique. The correlation of signatures between the test frame and \textit{reference frame} is calculated and classified by using a pre-determined optimal threshold. The spliced frame can be detected and localized if the correlation with the reference frame is less than the threshold. A refining step is further considered to reduce the detection and localization errors. The performance of the proposed scheme is tested using two data sets: (1) TIMIT database; (2) Real world audio recordings. Experimental results show that the proposed system can successfully identify the spliced frames in uncompressed audio recordings. Robustness of the proposed method is also tested for lossy compression attach, e.g., MP3 compression. In addition, performance of the proposed method is also compared against existing state-of-the-art work\cite{pan:2012}. The detection performance is also superior to the existing methods.

The rest of the paper is organized as follows. A brief overview of the state-of-art audio forensics is provided in Section \ref{sec:Pre_work}. The audio signal model and the environmental signature estimation algorithm are introduced in Section \ref{sec:Audio_Sig_Model_and_Feat_Est}. Applications of the environmental signature for audio authentication and splicing detection are elaborated in Section \ref{sec:Ap_AF}. Experimental setup, results, and performance analysis are provided in Section \ref{sec:results}. Finally, the conclusion is drawn in Section \ref{sec:diss} along with the discussion of future research directions.

\section{Related Works}\label{sec:Pre_work}

The research on audio/speech forensics dates back to the 1960s, when the U.S. Federal Bureau of Investigation has conducted an examination of audio recordings for speech intelligibility enhancement and authentication\cite{koenig:2007}. In those days, most forensic experts worked extensively with analog magnetic tape recordings to authenticate the integrity of the recording \cite{boss:2010,Begault:2005} by assessing the analog recorder fingerprints, such as length of the tape, recording continuity, head switching transients, mechanical splices, overdubbing signatures and etc. Authentication becomes more complicated and challenging for digital recordings as the trace of tampering is much more difficult to detect than mechanical splices or overdubbing signatures in the analog tape. Over the past few decades, several efforts have been initiated to fill the rapidly growing gap between digital media manipulation technologies and digital media authentication tools.

For instance, the electric network frequency (ENF)-based methods \cite{Rodriguez:2010, grigoras:2009a,cooper:2008,brixen:2007,Bykhovsky:2013,Hajj-Ahmad:2013} utilize the random fluctuations in the power-line frequency caused by mismatches between the electrical system load and generation. This approach can be extended to recaptured audio recording detection\cite{Su:2013} and geolocation estimation\cite{Garg:2013}. However, the ENF-based approaches may not be applicable if well-designed audio equipment (e.g., professional microphones) or battery-operated devices (e.g., smartphones) are used to capture the recordings. Although, the research on the source of ENF in battery-powered digital recordings is initiated\cite{Jidong:2013}, its application in audio forensics is still untouched.

On the other hand, statistical pattern recognition techniques were proposed to identify recording locations \cite{kraetzer:2007,oermann:2005,buchholz:2010, Malkin:2005,Chu:2006, Eronen:2006} and acquisition devices \cite{ garcia:2010, kraetzer:2009,Panagakis:2012,Kraetzer:2011,Kraetzer:2012,malik:2012c,Panagakis:2013}. For instance, techniques based on time/frequency-domain analysis (or mixed domains analysis) \cite{yang:2008,yang:2009,yang:2010,Korycki:2013} have been proposed to identify the trace of double compression attack in MP3 files. A framework based on frequency-domain statistical analysis has also been proposed by Grigoras \cite{grigoras:2010} to detect traces of audio (re)compression and to discriminate among different audio compression algorithms. Similarly, Liu \textit{et al.}~\cite{Liu:2010} have also proposed a statistical-learning-based method to detect traces of double compression. The performance of their method deteriorates for low-to-high bit rate transcoding. Qiao \textit{et al.} \cite{Qiao:2010} address this issue by considering non-zero de-quantized Modified Discrete Cosine Transform (MDCT) coefficients for their statistical machine learning method and an improved scheme has been proposed recently\cite{Qiao:2013}.  Brixen in \cite{brixen:2009} has proposed a time-domain method based on acoustic reverberation estimated from digital audio recordings of mobile phone calls for crime scene identification. Similar schemes using reverberation time for authentication have been studied by Malik\cite{malik:2010,Malik:2013a}. However, these methods are limited by their low accuracy and high complexity in model training and testing. Additionally, their robustness against lossy compression is unknown. In \cite{Hong:2013a,malik:2010,malik:2012a, malik:2012b, malik:2012c,zhao:2012}, model-driven approaches to estimate acoustic reverberation signatures for automatic acoustic environment identification and forgery detection where the detection accuracy and robustness against MP3 compression attacks are improved.

Recently, audio splicing detection has attracted a number of research interests. However, all the above methods were not primarily designed for audio splicing detection and their applicability for audio splicing detection is unclear. A method based on higher-order time-differences and correlation analysis has been proposed by Cooper \cite{cooper:2010} to detect traces of ``\emph{butt-splicing}'' in digital recordings. Another audio splicing detection scheme is based on the analysis of high-order singularity of wavelet coefficients\cite{Chen:2013}, where singular points detected by wavelet analysis are classified as forged. Experimental results show that the best detection rate of this scheme on WAV format audios is less than 94\% and decreases significantly with the reduction of sampling rate. Pan \textit{et al}. \cite{pan:2012} have proposed a time-domain method based on higher-order statistics to detect traces of splicing. The proposed method uses differences of the local noise levels in an audio signal for splice detection. It works well on simulated spliced audio, which is generated by assembling a pure clean audio and a noise audio, however, its performance in practical applications is unknown. In all the aforementioned schemes, it is assumed that the raw audio recording is available, and its performance might degrade significantly when lossy audio compression algorithm is used. In this paper, a robust feature, that is, the magnitude of acoustic impulse response, is proposed for audio splicing detection and localization. Experimental results demonstrate that the proposed scheme outperforms the state-of-art works described above.

\section{Audio Signal Modeling and Signature Estimation}\label{sec:Audio_Sig_Model_and_Feat_Est}

\subsection{Audio Signal Model}\label{sec:ASM}
\begin{figure}[htbp]
  \centering
  \epsfig{figure=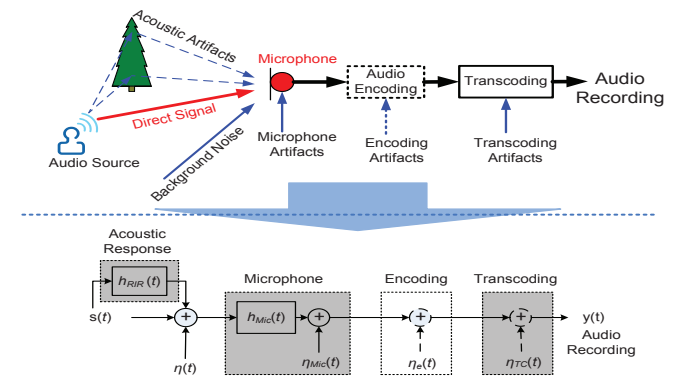,width=3.4in, height=2.0in}
  \caption{A simplified digital audio recording system diagram.}
  \label{fig:BlockDigram}
\end{figure}

In audio recording, the observed audio/speech will be contaminated due to the propagation through the acoustic channel between the speaker and the microphone. The acoustic channel can be defined as the combined effects of the acoustic environment, the locations of microphone and the sound source, the characteristics of the microphone and associated audio capturing equipment. In this paper, we consider the model of audio recording system in\cite{Hong:2013a}. For the beginning, let the $s(t)$, $y(t)$, $\eta(t)$ be the direct speech signal, digital, recording and background noise, respectively. The general model of recording system is illustrated in Fig. \ref{fig:BlockDigram}. In this paper, we assume that the the microphone noise, $\eta_{Mic}(t)$, and transcoding distortion, $\eta_{TC}(t)$, are negligible because of that high quality commercial microphone is used and the recordings are saved as raw format. One kind of transcoding distortion (lossy compression distortion) will be considered in the experiment later. With these assumptions, the model in Fig. \ref{fig:BlockDigram} can be simplified as,

%we illustrated the process of audio recording system. Consider a digital audio recording signal $x(t)$, which is a combination of several components such as, direct speech signal $s(t)$ (also called clean speech), acoustic environment distortion (consisting of \emph{reverberant signal} and \emph{background noise} $\eta(t)$), microphone distortion $\eta_{Mic}(t)$, encoding distortion $\eta_{e}(t)$, and transcoding distortion $\eta_{TC}(t)$. The simplified model for digital audio recording is shown in Fig. \ref{fig:BlockDigram}.

\begin{equation}
\begin{array}{rcl}
 y(t) & \approx & {h}_{RIR}(t)*h_{Mic}(t)*s(t) +{\eta}(t) \\
      & \approx & \dot{h}_{RIR}(t)*s(t) + {\eta}(t)
 \end{array}
 \label{eq:model}
\end{equation}
%The combined effects of the direct and reflected signals and the background (or ambient) noise at the input of the microphone can be expressed as:
%
%\begin{equation}
%  x(t) = s(t)*h_{RIR}(t)+ \eta(t)
%  \label{eq:sys1}
%\end{equation}
\noindent
where $*$ represents the convolution operation; $h_{RIR}$ denotes the Room Impulse Response (RIR) (or Acoustic Channel Response (ACR)) and $\dot{h}_{RIR}(t) = h_{Mic}(t)*h_{RIR}(t)$.
%If $h_{Mic}(t)$ denotes the microphone impulse response, then the digital audio recording, $y(t)$, can be expressed as:
%
%\begin{equation}
% y(t) = h_{Mic}(t)* x(t)+ \eta_{Mic}(t)+\eta_{TC}(t)
% \label{eq:sys2}
%\end{equation}

%We also assume that the impulse response of microphone, $h_{Mic}(t)$, is flat and short-time-invariant. With these assumptions,
%Eq.(\ref{eq:sys2}) can be reduced to
%
%\begin{equation}
% y(t) \approx \dot{h}_{RIR}(t)*s(t) + \dot{\eta}(t)
% \label{eq:sys3}
%\end{equation}
%
%\noindent
%where $\dot{\eta}(t) = h_{Mic}(t)*\eta(t)$, and $\dot{h}_{RIR}(t) = h_{Mic}(t)*h_{RIR}(t)$.

As a remark note that, $h_{RIR}(t)$ is uniquely determined by the acoustic propagation channel. $h_{Mic}(t)$ characterizes the property of the microphone used to capture the audio. Compared to $h_{RIR}(t)$, $h_{Mic}(t)$ is flatter, less dynamic and weaker for high quality microphone. It has been proved that $h_{Mic}(t)$ is also useful to authenticate the integrity of audio\cite{malik:2012c}. Therefore, we consider the combined effect of $h_{Mic}(t)$ and ${h}_{RIR}(t)$, and estimate them simultaneously. In the remaining of this paper, we shall refer to $\dot{h}_{RIR}(t)$ as \emph{environmental signature}, as ${h}_{RIR}(t)$ is predominant. Since the environment signature uniquely specifies the acoustic recording system, it can be used to authenticate the integrity of captured audio. The blind estimation algorithm of environment signature in the existing literature will be elaborated in Section \ref{sec:EnviSign_est}, and our proposed authentication and detection algorithm will be introduce in Section \ref{sec:Ap_AF}.
%For simplification, we use ${h}_{RIR}(t)$ instead of $\dot{h}_{RIR}(t)$ to represent the combined effect of impulse responses. In this paper, we are trying to use the environmental signature to uniquely authenticate the integrity of captured audios. In the following section, we shall first introduce the blind estimation algorithm for the environmental signature\cite{Nikolay:2013}.

\subsection{Environmental Signature Estimation}\label{sec:EnviSign_est}

Blind dereverberation is a process of separating reverberant and dry signals from a single channel audio recording without exploiting the knowledge of $\dot{h}_{RIR}(t)$. Dereverberation methods can be divided into the following categories: (i) frequency domain methods \cite{Nikolay:2013,Soulodre:2010,Wolfel2:2009,Sehr:2010}, (ii) time domain methods \cite{Bengt:2012,Nakatani:2010}, (iii) joint time-frequency domain methods \cite{Doclo:2001}. Blind dereverberation has found a wide range of applications ranging from speech enhancement\cite{Soulodre:2010,Wolfel2:2009} to distant speech recognition~\cite{Sehr:2010}, acoustic environment identification \cite{malik:2010, malik:2012a,zhao:2012,Hong:2013a}, audio forensics \cite{sohaib:2010, malik:2010}, etc. In our previous works \cite{malik:2010, malik:2012a,zhao:2012,Hong:2013a}, it was also experimentally verified that reverberation is quite effective for environment identification \cite{malik:2012a,zhao:2012,Hong:2013a} and audio forensics \cite{sohaib:2010, malik:2010} applications. In this paper, we shall use the frequency domain method to estimate the RIR. Specifically, we first introduce the environment signature which presents the log-spectrum magnitude of RIR contaminated by ambient noise, and then, elaborate on the corresponding estimation algorithm\cite{Nikolay:2013}. It can be archived by three steps approach as follows.

\textbf{Step 1:} Signature Expression\\
The Short Time Fourier Transform (STFT) representation of (\ref{eq:model}) can be expressed as:

\begin{equation}
  Y(k,l) = S(k,l) \times H(k,l) + V(k,l)
  \label{eq:freq_exp}
\end{equation}

\noindent
where $Y(k,l)$, $S(k,l)$, $H(k,l)$ and $V(k,l)$ are the STFT coefficients (in the $k^{th}$ frequency bin and the $l^{th}$ frame) of $y(t)$, $s(t)$, $\dot{h}_{RIR}(t)$ and $\eta(t)$, respectively. It is important to mention that (\ref{eq:freq_exp}) is valid if and only if the frame length of the STFT is larger than the duration of RIR, $h_{RIR}(t)$. Generally, acoustic channel varies much more slowly than the speech and, therefore, it is reasonable to assume that $|H(k,l)|$ does not vary significantly with $l$. Power spectrum of the noisy observation $y(t)$ is expressed as

\begin{equation}
  \begin{array}{rcl}
    |Y(k,l)|^2 &=& |S(k,l)|^2 \times |H(k,l)|^2 + |V(k,l)|^2 + \\
    & & 2\times |S(k,l)| \times |H(k,l)| \times |V(k,l)|\times \cos\theta \\
  \end{array}
  \label{eq:power_spectrum}
\end{equation}

\noindent
where $\theta = \angle S(k,l) + \angle H(k,l) - \angle V(k,l)$. Dividing both sides of (\ref{eq:power_spectrum}) by $|S(k,l)|^2 \times |H(k,l)|^2$ and transforming the resulting expression using logarithm operation results in:

\begin{equation}
  \log |Y(k,l)| - \log |S(k,l)| = \log |H(k,l)| + \log \varepsilon(k,l)
  \label{eq:log_spectrum}
\end{equation}

\noindent
where

\begin{equation}
  \varepsilon = \sqrt{\left( \xi^{-1} + 2\xi^{-0.5}\cos\theta + 1 \right)}\;,
\end{equation}

\noindent
and

\begin{equation}
  \xi = \dfrac{|S(k,l)|^2 \times |H(k,l)|^2}{|V(k,l)|^2},
\end{equation}

Gaubitch \textit{et al.} in \cite{Nikolay:2013} has shown the way to extract $\log |H(k,l)|$ and $\log\varepsilon(k,l)$ from the noisy observation $|Y(k,l)|$. However, in the application of audio forensics, the background noise provides useful information and its effectiveness has been experimentally verified in \cite{sohaib:2010,zhao:2012} and \cite{Hong:2013a}. Thus, we define the environmental signature $\underline{H}$ as follows

\begin{equation}
  \underline{H}(k) = \dfrac{1}{L}\sum_{l=1}^{L}\left( \log \left|Y(k,l)| - \log |S(k,l)\right| \right)
  \label{eq:est_RIR}
\end{equation}

\noindent
%\begin{equation}
%  \hat{H}(k) = \log |H(k)| + \log \varepsilon(k)
%\end{equation}
%
%\noindent
Note that in the definition of environmental signature (\ref{eq:est_RIR}), $Y(k,l)$ is the spectrum of the observation audio signal, and $S(k,l)$ is the spectrum of the clean audio signal, which can be estimated as follows.
%In practice, $S(k,l)$ is not available and can be approximated via a statistical model, such as the \emph{Gaussian Mixture Model (GMM)}. In the next section, the approximation of the clean log-spectrum, $\hat{S}(k,l)$, is discussed.

\textbf{Step 2:} {Clean Log-Spectrum Estimation}\\
In practice, $S(k,l)$ is not available and can be approximated via the \emph{Gaussian Mixture Model (GMM)} \cite{Bishop:2006}. Given a data set consisting of clean speeches with various speakers and contents, each sample, $s(n)$, is divided into overlapping windowed frames and proceeded by the STFT to obtain $S(k,l)$. Mean substraction is used to smooth the log-spectrum $S(k,l)$,

\begin{equation}
  S_{\log}(k,l) = \log |S(k,l)| - \dfrac{1}{K}\sum_{k=1}^{K}\log |S(k,l)|
\end{equation}

\noindent
where $K$ is the total number of frequency bins and $S_{\log}(k,l)$ represents the smoothed log-spectrum.

For each frame, the RASTA filtered Mel-Frequency Cepstral Coefficients (MFCC)\cite{Hynek:1994} (more robust than pure MFCC) are estimated, which results in an $N$-dimensional feature vector $\mathbf{mfcc_s(l)} = [mfcc_s(1,l),\dots, mfcc_s(N,l)]^T$, where $l$ denotes the $l^{th}$ audio frame. The target $M$-mixture GMM defined by a mean vector, $\mathbf{\mu}_m$, diagonal covariances matrix, $\mathbf{\Sigma}_m$, and weights vector, $\mathbf{\pi}_m$, of each mixture, is trained by using the MFCCs. The mixture probabilities are expressed as,

\begin{equation}
  \gamma_{l,m} = \dfrac{\mathbf{\pi}_m \mathcal{N}(\mathbf{mfcc_s(l)}|\mathbf{\mu}_m, \mathbf{\Sigma}_m)}{\sum_{i=1}^{M} \mathbf{\pi}_i \mathcal{N}(\mathbf{mfcc_s(l)}|\mathbf{\mu}_i, \mathbf{\Sigma}_i)}
  \label{eq:post_prob}
\end{equation}

\noindent
where $\mathcal{N}(\mathbf{mfcc_s(l)}|\mathbf{\mu}_m, \mathbf{\Sigma}_m)$ represents the multivariate Gaussian distribution.

The log-spectrum of clean speech can be obtained by weighted averaging over all frames as:

\begin{equation}
  \bar{S}_{\log}(k) = \dfrac{\sum_{l=1}^{L}\gamma_{l,m}\times S_{\log}(k,l)}{\sum_{l=1}^{L}\gamma_{l,m}}
  \label{eq:clean_spectrum}
\end{equation}

\textbf{Step 3:} Signature Estimation\\
Then, the clean speech model of (\ref{eq:clean_spectrum}) is now used to estimate the acoustic distortion in (\ref{eq:est_RIR}). More specifically, the following steps are involved in estimating acoustic distortion given in (\ref{eq:est_RIR}): (i) a feature vector, $\mathbf{mfcc_y(l)} = [mfcc_y(1,l),\dots, mfcc_y(N,l)]^T$, of observation $y(t)$ is obtained by applying RASTA filtering on estimate MFCC feature vector, (ii) the posterior probability of $\mathbf{mfcc_y(l)} $ is calculated by substituting filtered feature vector into (\ref{eq:post_prob}), which results in $\bar{\gamma}_{l,m}$ for each mixture $m=1,2,\dots,M$. With the availability of posterior probability, spectrum of the direct signal of the $l^{th}$ frame can be expressed as:

\begin{equation}
  \hat{S}(k,l) = \exp\left[{\sum_{m=1}^{M}\bar{\gamma}_{l,m} \times \bar{S}_{\log}(k) }\right]
  \label{eq:est_spectrum}
\end{equation}

Substituting (\ref{eq:est_spectrum}) into (\ref{eq:est_RIR}), the estimted signature can be expressed as:
\begin{equation}
  \underline{\hat{H}}(k) = \dfrac{1}{L}\sum_{i=1}^{L}\left[ \log |Y(k,l)| - \log\left(\left|e^{\sum_{m=1}^{M}\bar{\gamma}_{l,m} \times \bar{S}_{\log}(k)}\right|\right) \right]
 \end{equation}

As a remark note that the accuracy of the estimated RIR depends on the error between estimated spectrum, $\widehat{S}(k,l)$, and the true spectrum, $S(k,l)$.

\section{Applications to Audio Forensics}\label{sec:Ap_AF}

A novel audio forensics framework based on the acoustic impulse response is proposed in this section. Specifically, we consider two application scenarios of audio forensics: (i) \emph{audio source authentication} which determines whether the query audio/speech is captured in a specific environment or location as claimed; and (ii) \emph{splicing detection} which determines whether the query audio/speech is original or assembled using multiple samples recorded in different environments and detects splicing locations (in case of forged audio).

\subsection{Audio Source Authentication}\label{sec:Audio_Source_Auth}

\begin{figure}[thb]
  \centering
  \includegraphics[width=3.4in,height = 1.5in]{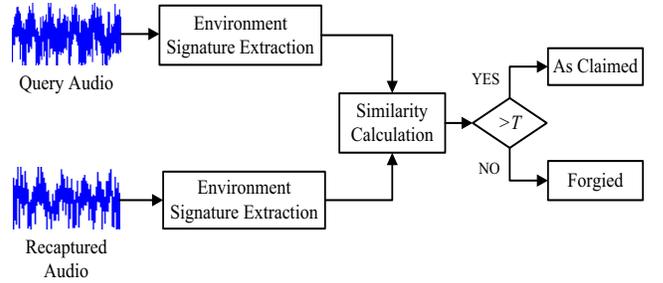}\\
  \caption{Block diagram of the proposed audio source authentication method.}
  \label{fig:proc_flowchart}
\end{figure}

In practice, the crime scene is usually secured and kept intact for possible further investigation. Thus, it is possible to rebuild the acoustic environmental setting, including the location of microphone and sound source, furniture arrangement, etc. Hence, it is assumed that the forensic expert can capture reference acoustic environment signature, denoted as $\underline{\hat{H}}_{r}$,  and compare it to the signature $\underline{\hat{H}}_q$ extracted from the query audio recording. As illustrated in Fig. \ref{fig:proc_flowchart}, the proposed authentication scheme is elaborated below:
\begin{enumerate}
  \item Estimate the RIRs, $\underline{\hat{H}}_q$ and $\underline{\hat{H}}_{r}$, from the query audio and reference audio according to the method described in Section \ref{sec:EnviSign_est}, respectively,
  \item Calculate the normalized cross-correlation coefficient $\rho$ between $\underline{\hat{H}}_q$ and  $\underline{\hat{H}}_{r}$ according to (\ref{eq:NCC}),

      \begin{equation}
        \begin{array}{rcl}
            \rho & =& \mathcal{NCC}(\underline{\hat{H}}_q,\underline{\hat{H}}_{r})  \\ [0.1in]
            & = & \dfrac{\sum\limits_{k=1}^{K}(\underline{\hat{H}}_q(k) - \mu_{q})(\underline{\hat{H}}_{r}(k) - \mu_{r})}{\sqrt{\sum\limits_{k=1}^{K}(\underline{\hat{H}}_q(k) - \mu_{q})^2}\sqrt{\sum\limits_{k=1}^{K}(\underline{\hat{H}}_{r}(k) - \mu_{r})^2}}
        \end{array}
        \label{eq:NCC}
      \end{equation}
    \noindent
    where $\mu_{q}= \frac{1}{K}\sum_{k=1}^{K} \underline{\hat{H}}_q(k)$, $\mu_{r}=\frac{1}{K}\sum_{k=1}^{K} \underline{\hat{H}}_{r}(k)$.

  \item Compare the cross-correlation coefficient with an threshold $T$ to determine the authentication of the query audio. Specifically, the following hypothesis testing rule is used,
      \begin{equation}
      \begin{array}{lc}
        \rho < T: &\text{Forged} \\
        \rho \geq T: & \text{As Claimed}
      \end{array}
      \label{eq:decision}
      \end{equation}
\end{enumerate}

The performance of the above authentication scheme depends on the threshold $T$. In the following part, an criteria is proposed to derive the optimal threshold.

\subsubsection{Formulation of Optimal Threshold}

Supposing that $f_e(\rho)$ represents the distribution of $\rho$, which is the correlation coefficient between the reference audio and the query audio captured in identical environment. In this case, $\underline{\hat{H}}_q$ and  $\underline{\hat{H}}_{r}$ are highly positive correlated. Similarly, $f_g(\rho)$ represents the distribution of $\rho$, which is the correlation coefficient between the reference audio and the query audio captured in different environments. In this case, $\underline{\hat{H}}_q$ and  $\underline{\hat{H}}_{r}$ are expected to be independent. Due to the inaccurate estimation of signatures and the potential noise, the resulted correlation coefficients will diverge from the true values. The statistical distributions of $\rho_i$s will be used to determine the optimal threshold. The binary hypothesis testing based on an optimal threshold, $T$ in (\ref{eq:decision}), contributes to the following two types of errors:
\begin{itemize}
  \item \textbf{Type \uppercase\expandafter{\romannumeral1} Error:} Labeling an authentic audio as \emph{forged}. The probability of Type \uppercase\expandafter{\romannumeral1} Error is called \emph{False Positive Rate (FPR)}, which is defined as,
      \begin{eqnarray}
        F_{PR}(T) &= &\int_{-\infty}^{T} f_e(\rho)\mathrm{d}x \nonumber \\ [0.1in]
                &= & 1-\mathcal{F}_e(T)
        \label{eq:evd_FPR}
      \end{eqnarray}
      where, $\mathcal{F}_e$ is the \emph{Cumulative Distribution Function (CDF)} of $f_e$.
  \item \textbf{Type \uppercase\expandafter{\romannumeral2} Error:} Labeling a forged audio as \emph{authentic}. The probability of Type \uppercase\expandafter{\romannumeral2} Error is called \emph{False Negative Rate (FNR)}, which is defined as,
      \begin{eqnarray}
        F_{NR}(T) & = & \int_{T}^{\infty}f_g(\rho)\mathrm{d}x \nonumber \\ [0.1in]
                & = & 1 - \mathcal{F}_g(T)
        \label{eq:ggd_FNR}
      \end{eqnarray}
      where, $\mathcal{F}_g$ is the \emph{(CDF)} of $f_g$.
\end{itemize}

Then, the optimal decision boundary $T$ can be determined by minimizing the combined effect of these errors as follows,

\begin{equation}
  T:=\arg \min\limits_{T\in [-1,1]}\left[{\lambda \times F_{PR}(T) + (1 - \lambda)\times F_{NR}(T)}\right]
  \label{eq:T_est}
\end{equation}

\noindent
where $\lambda\in [0,1]$ is the control factor chosen by the forensic expert according to the practical applications. In the following section, we are trying to fit $\rho$ using statistical models.

%
%\subsubsection{Approximated Closed-Form Solution for Threshold}
%In order to determine the optimal threshold, exhaustive search technique can be used by setting a small positive step from -1 to 1. It will result in high computing complexity. However, in the practical applications, the threshold should be chosen according the maximal allowed FPR $\tau$. This is very common, as the FPR of the forensic system is usually required to be kept below a specific rate $\tau$. This requires the closed-form of $f_e(\rho_i)$ and $f_g(\rho_i)$. If the closed-form of $f_e(\rho_i)$ and $f_g(\rho_i)$ are available, the optimal threshold regarding to the specific FPR $\tau$ can be obtained according to (\ref{eq:evd_FPR}) as follows,
%
%\begin{equation}
%  T = \mathcal{F}_e^{-1}(1 - \tau)
%  \label{eq:T_est_by_tau}
%\end{equation}
%\noindent
%where $\mathcal{F}_e^{-1}$ represents the inverse function of $\mathcal{F}_e$. Substituting (\ref{eq:T_est_by_tau}) into (\ref{eq:ggd_FNR}) results in the FNR as follows
%
%\begin{equation}
%  FNR_T = 1 - \mathcal{F}_g(\mathcal{F}_e^{-1}(1 - \tau))
%  \label{eq:FNR_tau}
%\end{equation}
%
%In this following section, we are trying to fit $\rho_i$ using statistical models and derive the closed-form of (\ref{eq:T_est_by_tau}).

Deriving the optimal threshold in (\ref{eq:T_est}) requires the prior knowledge of $f_e(\rho)$ and $f_g(\rho)$. One intuitive solution is using the empirical distributions (histograms) of $\rho$. However, this strategy results in high computing complexity of determining the optimal threshold in (\ref{eq:T_est}). The anther disadvantage is that the closed-form of the threshold can not be determined according to specified FPR or FNR. In the next section, we fit the distributions using statistical models and elaborate the estimation of the model parameters, which can be used to calculate the threshold.

\subsubsection{Distribution Modeling and Parameters Estimation}

\begin{figure}[thbp]
  \centering
  \includegraphics[width=3.5in,height=2.5in]{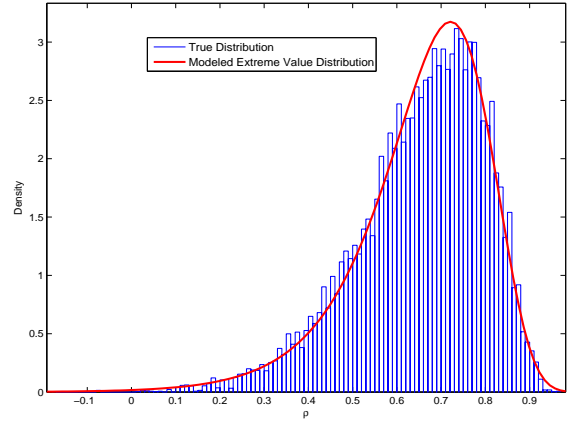}\\
  \caption{Plots of the true distribution of $\rho$ and the extreme value distribution with estimated parameters $\hat{\mu}_e = 0.72$ and $\hat{\delta}_e = 0.11$.}
  \label{fig:evd_est_example}
\end{figure}

To model distribution of $\rho$, we consider the same acoustic environment case first. That is, for identical acoustic environments, correlation coefficient, $\rho$, is very close to the extreme value (maximum value) of interval $[-1, 1]$, e.g.  $\rho\rightarrow 1$. We experimentally find that the extreme value distribution is a good model in this case due to its simplification and low fitting error(shown in Fig. \ref{fig:evd_est_example} and Section\ref{sec:Res_Synth_Data}). The \emph{Probability Density Function (PDF)} of \emph{extreme value distribution} is expressed as \cite{Kotz:2000},

\begin{equation}
  f_{e}(\rho |\mu_{e},\delta_{e}) = \delta_{e}^{-1}\exp\left[\frac{\rho-\mu_{e}}{\delta_{e}}-\exp\left( \frac{\rho-\mu_{e}}{\delta_{e}}\right)  \right]
  \label{eq:EVD}
\end{equation}

\noindent
where $\mu_{e}$ and $\delta_{e}$ represent the location and scale parameter, respectively.  The parameters, $\mu_e$ and $\delta_e$, can be estimated via maximum likelihood estimation\cite{Eddy:1997}. The likelihood of drawing $N$ samples $\rho_i$ from an extreme value distribution with parameters $\mu_e$ and $\delta_e$ can be expressed as,

\begin{equation}
  P(\rho_1,\dots,\rho_N|\mu_e,\delta_e) =\prod_{i=1}^{N}\delta_e^{-1}\exp\left[\underline{\rho}_i-\exp\left( \underline{\rho}_i\right)  \right]
  \label{eq:evd_MLD}
\end{equation}

\noindent
and
\begin{equation}
  \underline{\rho}_i=\frac{\rho_i-\mu_{e}}{\delta_{e}}
\end{equation}

\noindent

The maximum likelihood estimation of  $\hat{\mu}_e$ and $\hat{\delta_e}$ can be expressed as follows,

\begin{equation}
  \hat{\mu}_e = \delta_e\log\left[\frac{1}{N}\sum_{i=1}^{N}\exp\left({\rho_i\hat{\delta}_e^{-1}}\right)\right]
  \label{eq:est_muE}
\end{equation}

\begin{equation}
  -\hat{\delta}_e - \frac{1}{N}\sum_{i=1}^{N}\rho_i + \frac{\sum_{i=1}^{N}\rho_{i}\exp\left[{\rho_i\hat{\delta}_e^{-1}}\right]}{\sum_{i=1}^{N}\exp\left[{\rho_i\hat{\delta}_e^{-1}}\right]} = 0
  \label{eq:evd_est_delta}
\end{equation}

Numerical methods such as gradient descent can be used to solve (\ref{eq:evd_est_delta}). Fig. \ref{fig:evd_est_example} shows the plots of the true distributions of $\rho$, for the case of that the query audio and the recaptured audio are in the same environment, and the extreme value distribution with estimated parameters $\hat{\mu}_e = 0.72$ and $\hat{\delta}_e = 0.11$. It can be observed from Fig. \ref{fig:evd_est_example} that, the estimated distribution is skewed left, and therefore, the extreme value distribution can model the true distribution reasonably well.

\begin{figure}[thbp]
  \centering
  \includegraphics[width=3.5in,height = 2.5in]{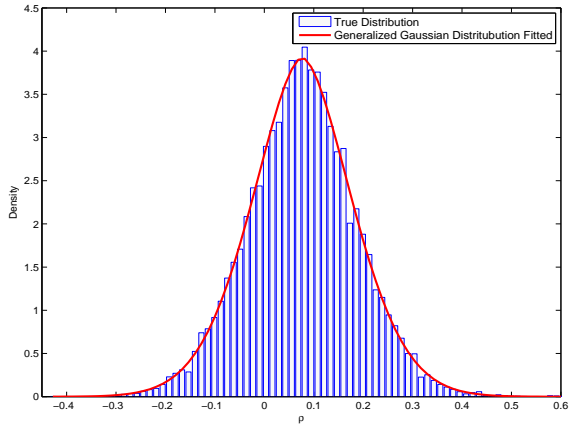}\\
  \caption{Plot of the true distribution of $\rho$ and \emph{generalized Gaussian distribution} with estimated parameters $\hat{\mu}_g = 0.077$, $\hat{\alpha}_g = 0.14$ and $\hat{\beta}_g = 1.74$.}
  \label{fig:ggd_est_example}
\end{figure}

Similarly, correlation coefficient $\rho$ between $\underline{\hat{H}}_q(k)$ and $\underline{\hat{H}}_{r}(k)$ estimated from the query audio and the recaptured audio which were made in two different environments is expected to be close to zero. In this case, correlation coefficient $\rho$ is expected to obey a \emph{heavy-tailed} distribution. To this end, the distribution of $\rho(i),i=1,\dots,M$ is modeled using the \emph{Generalized Gaussian} \cite{Dominguez:2001} distribution. Motivation behind considering the generalized Gaussian model here is due to the fact that thicker tails will lead to more conservative error estimates, as demonstrated in Fig. \ref{fig:ggd_est_example}. The \emph{PDF} of \emph{Generalized Gaussian} distribution is given as follows,
\begin{equation}
  f_g(\rho |\alpha_g,\beta_g,\mu_g)=\frac{1}{2\alpha_g\Gamma(1/\beta_g)}\exp\left[{-\left(\frac{|\rho-\mu_g|}{\alpha_g}\right)^{\beta_g}}\right]
  \label{eq:GGM}
\end{equation}

\noindent
where $\alpha_g$, $\beta_g$, and $\mu_g$ represent the scale, shape and mean parameters, respectively.

The parameters can be estimated using the method of moments\cite{Dominguez:2001} as follows,

\begin{eqnarray}
    \hat{\mu}_g & = & \frac{1}{N}\sum_{i=1}^{N}\rho_i \\
    \hat{\beta}_g & = & G^{-1}(\hat{m}_1^2/\hat{m}_2) \\
    \hat{\alpha}_g & = & \hat{m}_1\frac{\Gamma(1/\hat{\beta}_g)}{\Gamma(2/\hat{\beta}_g)}
\end{eqnarray}

\noindent
and

\begin{eqnarray}
  \hat{m}_1 &=& \frac{1}{N}\sum_{i=1}^N|\rho_i-\hat{\mu}_g| \\
  \hat{m}_2 &=& \frac{1}{N}\sum_{i=1}^N|\rho_i-\hat{\mu}_g|^2 \\
  G(x) &=& \frac{\left[\Gamma(2/x)\right]^2}{\Gamma(1/x)\Gamma(3/x)}
\end{eqnarray}

\noindent
where $\Gamma(x)$ is the Gamma function.

Fig. \ref{fig:ggd_est_example} shows the plot of the distribution of normalized correlation coefficients between $\underline{\hat{H}}_q(k)$ and $\underline{\hat{H}}_{r}(k)$, which are estimated from audio samples captured in different environments. It can be observed from Fig. \ref{fig:ggd_est_example} that the distribution of $\rho$ fits reasonably well into the generalized Gaussian PDF. The goodness of fit is evaluated based on the \emph{Kullback-Leibler divergence} (KLD). The KLD between the true distribution and generalized Gaussian fitted distribution is equal to $0.24$ which indicates that the assumed generalized Gaussian distribution fits reasonably well with the assumed model. It can be observed from Figs. \ref{fig:evd_est_example} and \ref{fig:ggd_est_example} that both distributions are heavy-tailed  and therefore will overlap.

In some practical applications, the threshold should be selected to satisfy the \textit{maximum allowed FPR ($\tau$) requirement}. The optimal threshold regarding to the specific FPR $\tau$ can be obtained via

\begin{equation}
  T = \mathcal{F}_e^{-1}(1 - \tau)
  \label{eq:T_est_by_tau}
\end{equation}
\noindent
where $\mathcal{F}_e^{-1}$ represents the inverse function of $\mathcal{F}_e$. Subsequently, substituting (\ref{eq:T_est_by_tau}) into (\ref{eq:ggd_FNR}) results in the FNR as follows

\begin{equation}
  FNR_T = 1 - \mathcal{F}_g(\mathcal{F}_e^{-1}(1 - \tau))
  \label{eq:FNR_tau}
\end{equation}

With these fitted statistical model (\ref{eq:EVD}), (\ref{eq:T_est_by_tau}) can be deduced to
\begin{equation}
  T_{model} = \hat{\delta}_e \ln \left[ \ln \left( \dfrac{1}{1-\tau}\right) \right] + \hat{\mu}_e
\end{equation}

\subsection{Audio Splicing Detection and Localization}\label{Sec:Audio_Splicing_Det}

\begin{figure}[thbp]
  \centering
  \includegraphics[width=2.8in,height = 1in]{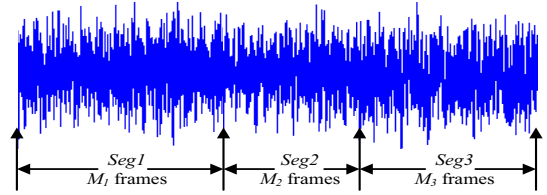}\\

  \caption{Plot of the spliced audio obtained assembled from three segments, $Seg1$, $Seg2$, and $Seg3$ consisting of $M_1$, $M_2$, and $M_3$ frames, respectively. Here, audio segments $Seg1$ and $Seg3$ are recorded in acoustic environment $A$ and audio segment $Seg2$ is recorded in environment $B$.}
  \label{fig:splciedaudio_example}
\end{figure}

%Audio splicing, referring to the creating an audio clip using multiple audio clips (may be recorded by different microphones at different environments), is the most common counterfeiting technology.
In this section, we will extend the method proposed in Section \ref{sec:Audio_Source_Auth} to audio splicing detection. To illustrate audio splicing process, an audio clip is assembled from three segments, $Seg1$, $Seg2$, and $Seg3$ consisting of $M_1$, $M_2$, and $M_3$ frames, respectively. Without loss of generality, it is assumed that audio segments $Seg1$ and $Seg3$ are recorded in acoustic environment $A$ and audio segment $Seg2$ is recorded in environment $B$. Fig.\ref{fig:splciedaudio_example} illustrates the temporal plot of the resulting audio. It can be observed from Fig. \ref{fig:splciedaudio_example} that splicing does not introduce any visual artifacts in the resulting audio.

\begin{figure}[thbp]
  \centering
  \includegraphics[width=3in,height = 3.5in]{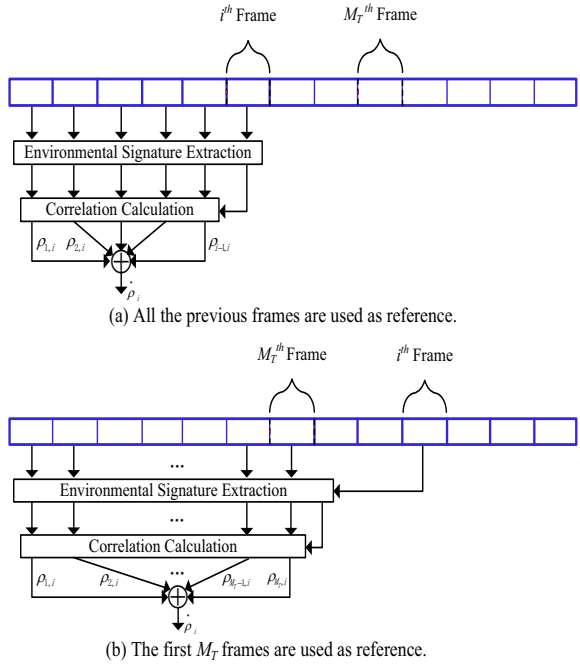}\\
  \caption{An flow chart of correlation calculation proposed for audio splicing detection. $\bigotimes$ and $\bigoplus$ represents the correlation calculation operator and arithmetical average operator, respectively.}
  \label{fig:splicing_det_ncc_cal_example}
\end{figure}

The difference between splicing detection and source authentication is that the reference audio is usually not available. Hence, in our proposed scheme, we shall use the RIRs extracted from the previous frames as reference, and compare it with the coming frames. Specifically, the proposed scheme can be divided into two phase: raw detection and refinement. The procedure of raw detection is elaborated below:

\begin{enumerate}
  \item Divide the query audio in to overlapped frames and estimate the RIR, $\underline{\hat{H}}_q^{i}$, from the $i^{th}$ frame according to the method in Section \ref{sec:EnviSign_est},
  \item Calculate the cross-correlation coefficient between the RIRs of the $i^{th}$ frame and the previous frames according to (\ref{eq:NCC_averaged}),
      \begin{equation}
          \dot{\rho}_i =  \begin{cases}
                    \dfrac{1}{i}\sum\limits_{j=1}^{i}\mathcal{NCC}(\underline{\hat{H}}_q^{(j)},\underline{\hat{H}}_{q}^{(i)}),&\text{if}\; \dot{\rho}_{i-1}>T, i \leq M_T \\
                    \dfrac{1}{M_T}\sum\limits_{j=1}^{M_T}\mathcal{NCC}(\underline{\hat{H}}_q^{(j)},\underline{\hat{H}}_{q}^{(i)}), & \text{otherwise}
                  \end{cases}
          \label{eq:NCC_averaged}
      \end{equation}

      \begin{equation}
          M_T = \min\limits_{i\in [1, M]}\{i\mid \dot{\rho}_i<T\} \text{ and } \dot{\rho}_1 = 1
      \end{equation}

     \noindent
     where $\underline{\hat{H}}_{q}^{(i)}$ is the estimated channel response of the $i^{th}$ frame from the query audio. The term $M$ is the length of query audio. $M_T$ is the rough estimation of $M_1$. Figs.  \ref{fig:splicing_det_ncc_cal_example} (a) and (b) show the correlation calculation methods expressed in (\ref{eq:NCC_averaged}) for $i\leq M_T$ and $i>M_T$, respectively.
  \item Identify the spliced frame of the query audio according to (\ref{eq:Blind_Detect_Binary_D})
     \begin{equation}
      \begin{array}{lc}
                   \dot{\rho}_i < T : & \text{Spliced Audio Frame} \\
                    \dot{\rho}_i \geq T: & \text{Original Audio Frame}
            \end{array}
      \label{eq:Blind_Detect_Binary_D}
      \end{equation}
\end{enumerate}

%In general, the expected value of $M_T = M_1$. However, it has been observed through experimental results that $M_T < M_1$ for most of the cases due to the estimation noise. Eq. (\ref{eq:NCC_averaged}) first calculates the cross-correlation coefficient between $\underline{\hat{H}}_q^{(1)}$ and $\underline{\hat{H}}_q^{(2)}$. If $\dot{\rho}_2 >T$, the second frame is authentic (come from the same location with the first frame). Then, it calculates the cross-correlation coefficients between $\underline{\hat{H}}_q^{(1)}$ and $\underline{\hat{H}}_q^{(3)}$, $\underline{\hat{H}}_q^{(2)}$ and $\underline{\hat{H}}_q^{(3)}$, updates the coefficients according to the smoothing method in (\ref{eq:NCC_averaged}) and moves forward to the next frame, and so on. The average-based smoothing technique used in (\ref{eq:NCC_averaged}) results in a stable estimation of RIRs, which will reduce the decision error of (\ref{eq:Blind_Detect_Binary_D}).

It should be noted that this framework for localizing forgery locations in audio requires {\it an accurate} estimate of the impulse response, $\underline{\hat{H}}_{q}(k)$. Any distortions in  $\underline{\hat{H}}_{q}(k)$, will result in a number of false alarms. However, due to the lack of substantial reference frames, the average-based smoothing technique (\ref{eq:NCC_averaged}) is adopted to reduce the detection errors. Eq. (\ref{eq:NCC_averaged}) first calculates the cross-correlation coefficient between $\underline{\hat{H}}_q^{(1)}$ and $\underline{\hat{H}}_q^{(2)}$. If $\dot{\rho}_2 >T$, the second frame is authentic (come from the same location with the first frame). Then, it calculates the cross-correlation coefficients between $\underline{\hat{H}}_q^{(1)}$ and $\underline{\hat{H}}_q^{(3)}$, $\underline{\hat{H}}_q^{(2)}$ and $\underline{\hat{H}}_q^{(3)}$, updates the coefficients according to the smoothing method in (\ref{eq:NCC_averaged}) and moves forward to the next frame, and so on. The average-based smoothing technique used in (\ref{eq:NCC_averaged}) results in a stable estimation of RIRs, which will reduce the decision error of (\ref{eq:Blind_Detect_Binary_D}).

Our experimental results show that the detection performance can be further improved by exploiting the correlation between adjacent frames. The motivation behind this is that tampering usually occurs in many consecutive frames instead of a few scattered frames. That means if neighborhoods of the current frame are forged, the current frame has a significatively high probability of being labeled as forged also. Based on this reasonable assumption, local correlation between the estimated signatures extracted from adjacent frames is used to make the forgery localization approach robust to estimation errors. The refinement procedure is summarized as follows:
\begin{enumerate}
  \item Detect the suspected frame using the raw detection method and set the label $p_k$ of the $i^{th}$ frame to 1 as follows,
      \begin{equation}
        p_k = 1, \text{ if } \dot{\rho}_i < T
      \end{equation}
  \item Refine the results using neighborhood similarity score as follows,
      \begin{equation}
      q_k = 1, \text{ if } \dfrac{\sum_{k=i-W/2}^{k=i+W/2}p_k-p_i}{W}>R_s
      \label{eq:sim_score}
      \end{equation}
      where, $W$ and $R_s$ represent the window size (e.g., $W$=3 or 5) and similarity score threshold (e.g., $R_s\in[0.7,0.9]$), respectively. $q_k=1$ indicates that the $k^{th}$ frame is detected as \textit{spliced}.
\end{enumerate}

Eq.(\ref{eq:sim_score}) means that, for the $i^{th}$ suspected frame, if the ratio of its adjacent frames being suspected exceeds $R_s$, the $i^{th}$ frame will be labeled as \textit{spliced}. It have been observed through experimentation that the refined algorithm significantly reduces the false alarm rates and improves detection performance. Section \ref{sec:results} provides performance analysis of the proposed method on both synthetic and real-world recordings.

%The pseudo-code of the proposed \textit{Robust Forgery Localization Algorithm} is shown in \textbf{ Algorithm \ref{Alg:Robust_Tam_Local_Alg}}.
%
%\begin{algorithm}[htp]
%\caption{Robust Forgery Localization Algorithm}
%\label{Alg:Robust_Tam_Local_Alg}
%\begin{algorithmic}[1]
%\REQUIRE \quad \\
%         $\dot{\rho}_i: \text{NCC between the $i^{th}$ frame and reference frame}$\\
%         $T: \text{Threshold}$\\
%         $W: \text{Number of frames for smoothing}$\\
%         $R_s: \text{Predefined Ratio}$\\
%         $N_f: \text{Total number of frames}$
%\ENSURE $q$: Indicate the locations of forged audio frames
%\STATE $i \leftarrow 1$
%\WHILE{$i<N_f$}
%\IF{$\dot{\rho}_i<T$}
%\STATE $p_i \leftarrow 1$ \label{alg:rough}
%\ENDIF
%\STATE $i\leftarrow i+1$
%\ENDWHILE
%\STATE $i\leftarrow W/2$
%\WHILE{$i<N_f-W/2$}
%\STATE $S \leftarrow 0$
%\FOR{$k=i-W/2$ to $i+W/2 \;\AND\; k\neq i$} \label{alg:coo1}
%\IF{$p_k=1$}
%\STATE $S\leftarrow S + 1$
%\ENDIF
%\ENDFOR
%\IF{$S/W>R_s$}
%\STATE $q_i \leftarrow 1 (Forged)$
%\ELSE
%\STATE $q_i \leftarrow 0 (Orignial)$
%\ENDIF \label{alg:coo2}
%\STATE $i \leftarrow i+1$
%\ENDWHILE
%\end{algorithmic}
%\end{algorithm}

\section{Performance Evaluation}\label{sec:results}
The effectiveness of the proposed methodology is evaluated using two data sets: synthetic data and real world human speech. Details for the data sets used, the experimental setup and the experimental results are provided below.

\subsection{Data Set and Experimental Setting}
Firstly, the speech data of TIMIT corpus\cite{TIMIT:1993} was used to train the clean log-spectrum and generate the synthetic data. TIMIT consists of 6300 sentences: ten sentences spoken by each of 438 male and 192 female speakers. The data set is divided into a training set (462 speakers) and a test set (168 speakers) with entirely different sentence contents apart from the dialect diagnostics. Each utterance is approximately 3 seconds long with sampling frequency $f_s = 16$ kHz. For training, the entire training set was processed using Hanning windowed frames (128ms for each) with overlapping rate of $50\%$. 12 RASTA-MFCCs of each frame was calculated and used to train the GMM with 512 mixtures. For test, the ten sentences of each speaker in the test set were concatenated to form one utterance with an approximate duration of 30s. Synthetic room response generated by source-image method\cite{Lehmann:2010} for a rectangular room is convolved with the test speech. Each time, all the parameters used for room response simulation were randomly selected to generate the RIRs of different environments. Table \ref{Tab:parameters} shows the candidate interval of parameters used.

\begin{table}[thbp]
  \centering
  \caption{The candidate range of each parameter used for room impulse response simulation}
  \begin{tabular}{|c|c|c|c|}
    \hline
    Parameter & Interval & Parameter & Interval \\ \hline
    \minitab[c]{Room Height \\$r_z$} & $[2.5, 3.5]$ & \minitab[c]{Mic Position\\ $m_x$} & $[0.5, r_x-0.5]$\\ \hline
    \minitab[c]{Room Width \\ $r_x$} & $[2.5, 5]$   & \minitab[c]{Mic Position\\ $m_y$} & $[0.5, r_y-0.5]$\\ \hline
    \minitab[c]{Room Length\\ $r_y$} & $[r_x, 10]$  & \minitab[c]{Sound Position\\ $s_z$} & $[0.4,r_z-0.8]$\\ \hline
    $T60$                            & $[0.1, 0.7]$  & \minitab[c]{Sound Position\\ $s_x$ } & $[0.3,r_x-0.9]$\\ \hline
    \minitab[c]{Mic Position\\ $m_z$}& $[0.5, 1.0]$ & \minitab[c]{Sound Position\\ $s_y$} & $[0.3,r_y-0.9]$ \\ \hline
  \end{tabular}
  \label{Tab:parameters}
\end{table}

For the second data set, real world data consisting of 60 speech recordings were used (the same data set used in\cite{Malik:2013a}). The speeches were recorded in four different environments: (1) outdoor; (2) small office (\mbox{$3.4$ m$\times 3.8$ m$\times 2.7$ m}, predominantly furnished with carpet and drywalls); (3) stairs (predominantly ceramic tiles and concrete walls); (4) restroom (\mbox{$5.2$ m$\times 4.3$ m$\times 2.7$ m}, predominantly furnished with ceramic tiles). In each recording environment, each of the three speakers (one male $S_1$, two females $S_2$ and $S_3$) read five different texts. The audio samples were captured by a commercial-grade external microphone mounted on a laptop computer using Audacity $2.0$ software. The audio was originally recorded with $44.1$ kHz sampling frequency, 16 bits/sample resolution, and then downsampled to $16$ kHz. The real world data set will be used to verify the effectiveness of the proposed scheme.

\subsection{Experimental Results}
\subsubsection{Results on Synthetic Data}\label{sec:Res_Synth_Data}
\begin{figure}[htbp]
  \centering
  \includegraphics[width=3.5in,height = 2.5in]{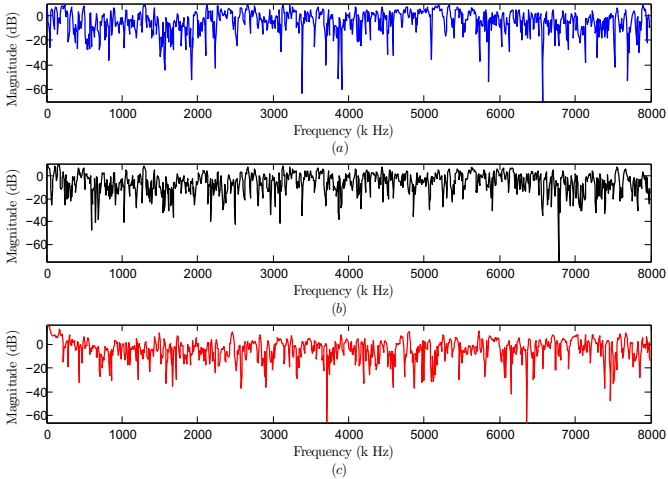}\\
  \caption{Magnitude (dB) plots of estimated $\underline{\hat{H}}$ from two different environments: (a) Estimated $\underline{\hat{H}}_{A1}$ from randomly tested speech 1; (b)  Estimated $\underline{\hat{H}}_{A2}$ from randomly tested speech 2, which is different from speech 1; (c)  Estimated $\underline{\hat{H}}_B$ from randomly tested speech 3, which is different from speeches 1 and 2.}
  \label{fig:sythetic_data_Est_RIR_examp}
\end{figure}

In this section, synthetically generated reverberant data were used to verify the effectiveness of the proposed algorithm. First, randomly selected parameters from Table \ref{Tab:parameters} were used to generate two different RIRs $H_A$ and $H_B$, which characterize two different environments $A$ and $B$. Both of them were convoluted with randomly selected speeches from the test set. Shown in Fig. \ref{fig:sythetic_data_Est_RIR_examp} are the magnitude plots of the estimated $\underline{\hat{H}}_A$ and $\underline{\hat{H}}_B$. It can be observed that from Fig. \ref{fig:sythetic_data_Est_RIR_examp} that the RIRs ($\underline{\hat{H}}_{A1}$ and $\underline{\hat{H}}_{A2}$) estimated from the speeches in environment $A$ are quite similar, and different from $\underline{\hat{H}}_{B}$, estimated from the speeches in environment $B$. The normalized correlation coefficient between $\underline{\hat{H}}_{A1}$ and $\underline{\hat{H}}_{A2}$ is equal to 0.81, which is significantly higher than that between $\underline{\hat{H}}_{A1}$ and $\underline{\hat{H}}_{B}$ (0.001). Similar results can be obtained from other tests. More results can be found in \cite{Nikolay:2013}.

\begin{figure}[htbp]
  \centering
  \includegraphics[width=3.5in,height = 2.2in]{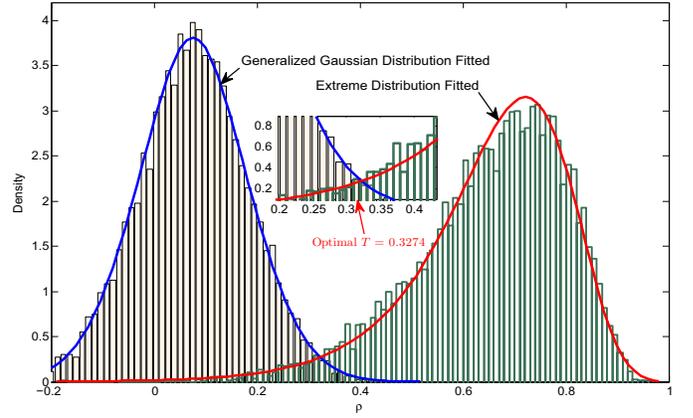}\\
  \caption{The true and the fitted distributions of $\rho_{A,i\rightarrow A,j}$ and $\rho_{A,i\rightarrow B,k}$ and the optimal decision threshold, $T=0.3274$}
  \label{fig:sythetic_data_Est_T_est}
\end{figure}

The optimal threshold $T$ can be determined from the synthetic data as follows. The randomly generated RIRs $H_A$ and $H_B$ were convolved with the test data set, respectively, resulting in two simulated synthetic data sets from two virtual environments $A$ and $B$. Then, the environmental signatures estimated from the synthetic speeches in environments $A$ and $B$ are represented as $\underline{\hat{H}}_{A,i}$ and $\underline{\hat{H}}_{B,i}$, respectively, where $0<i\leq N_t$, $N_t$ denotes total number of testing speeches. The distributions of intra-environment correlation coefficient, $\rho_{A,i\rightarrow A,j}=\mathcal{NCC}(\underline{\hat{H}}_{A,i},\underline{\hat{H}}_{A,j}), i\neq j$ and inter-environment correlation coefficient $\rho_{A,i\rightarrow B,k}=\mathcal{NCC}(\underline{\hat{H}}_{A,i},\underline{\hat{H}}_{B,k})$, are shown in Fig. \ref{fig:sythetic_data_Est_T_est}. As discussed in Section \ref{sec:Audio_Source_Auth}, $\rho_{A,i\rightarrow A,j}$ and $\rho_{A,i\rightarrow B,k}$ can be modeled as the extreme value distribution and the generalized Gaussian distribution, respectively. The optimal threshold $T$ for these distributions is determined as $0.3274$ by setting $\lambda = 0.5$, which results in the minimal overall error of $2.23\%$. It is important to mention that depending on the application at hand, $\lambda$ and the corresponding optimal decision threshold, $T$, can be selected.

\begin{figure}[htbp]
  \centering
  \includegraphics[width=2.5in,height = 1.5in]{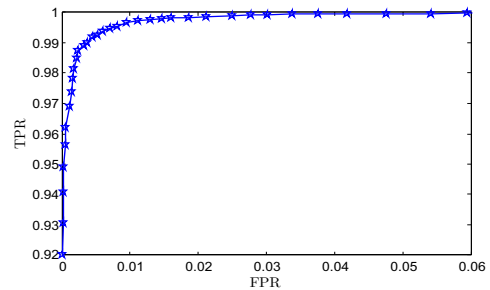}\\
  \caption{ROC curve of the proposed audio source authentication scheme.}
  \label{fig:sythetic_data_ROC}
\end{figure}

We have also evaluated performance of the proposed method in terms of the Receiver Operating Characteristic (ROC). Fig. \ref{fig:sythetic_data_ROC} shows the ROC curve of True Positive Rate (TPR,  which represents the probability that query audio is classified as \emph{forged}, when in fact it is) vs False Positive Rate (FPR, which represents the probability that query audio is labeled as \emph{forged}, when in fact it is not) on the test audio set. It can be observed from Fig. \ref{fig:sythetic_data_ROC} that, for FPR  $ > 3\%$, $TPR \rightarrow 1$. It indicates that the proposed audio source authentication scheme has almost perfect detection performance when the recaptured reference audios are available. In addition, the forensic analyst can recapture sufficient reference audio samples, which leads to more accurate and stable estimation of environment signature hence better detection performance.

\begin{figure}[htbp]
  \centering
  \includegraphics[width=3.5in, height = 2in]{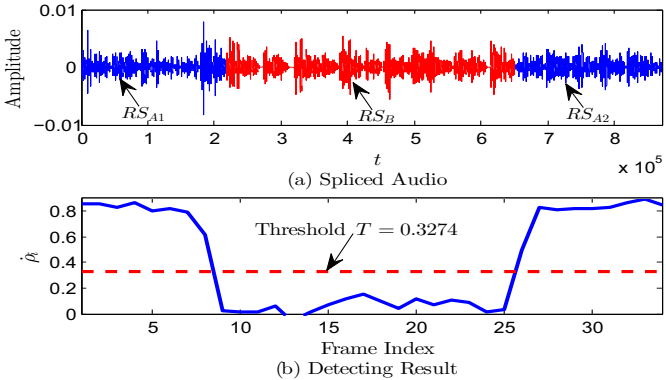}\\
  \caption{Forgery detection and localization results (with frame size = 3s).}
  \label{fig:sythetic_data_splicing_det_3s}
\end{figure}

Furthermore, we also evaluate the performance of the proposed splicing detection algorithm for synthetic data. Two randomly selected speeches from the test set were convolved with $H_A$ and $H_B$, resulting in two reverberant speeches $RS_A$ and $RS_B$. And $RS_B$ was inserted into a random location within $RS_A$. Fig. \ref{fig:sythetic_data_splicing_det_3s}(a) shows the plot (in time domain) of the resulting audio. It is barely possible to observe the trace of splicing by visual inspection, even for forensic experts. Fig. \ref{fig:sythetic_data_splicing_det_3s}(b) shows the detection output of the proposed algorithm listed in \textbf{Algorithm \ref{Alg:Robust_Tam_Local_Alg}} with a frame size of $3s$. It can be observed from Fig. \ref{fig:sythetic_data_splicing_det_3s} that the proposed forgery detection and localization algorithm has successfully detected and localized the inserted audio segment, $RS_B$ with $100\%$ accuracy.

\begin{figure}[htbp]
  \centering
  \includegraphics[width=3.5in,height = 2.2in]{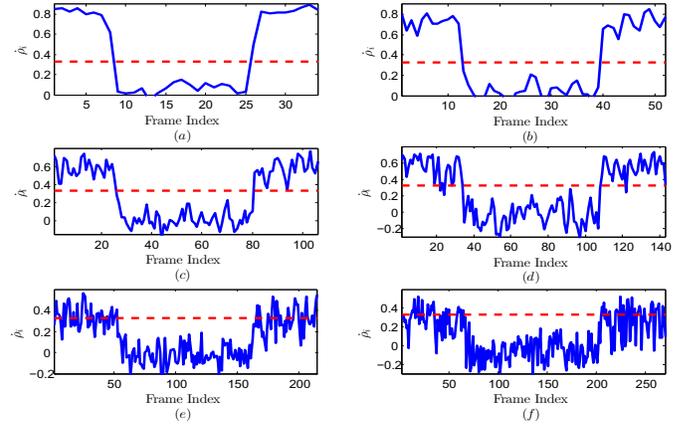}\\
  \caption{Detecting performance of the proposed scheme with various frame sizes: (a)3s, (b)2s, (c)1s, (d)750ms, (e)500ms, (f)400ms, Red line represents the threshold $T = 0.3274$ used to classify the suspected forged frame}
  \label{fig:synthetic_data_splicing_det_diff_frame}
\end{figure}

The impact of frame size on performance of the proposed scheme is investigated next. Fig. \ref{fig:synthetic_data_splicing_det_diff_frame} shows the detection performance of the proposed forgery detection and localization algorithm as a function of the frame size (used to estimate the environmental signature $\underline{\hat{H}}$). It is intuitive that a larger frame size would result in more accurate estimate of $\underline{\hat{H}}$, hence higher detection accuracy. Forgery localization, on the other hand, has the localization granularity ($\approx$ half of frame size) and thus requires a smaller frame size. In other words, decreasing frame size leads to a better localization resolution, but less accurate estimate of environment signature, which may reduce detection accuracy.  It can be observed from Fig. \ref{fig:synthetic_data_splicing_det_diff_frame} that for frame size $> 1$, the splicing parts can be identified and localized with $100\%$ accuracy. Subsequently, for frame sizes between $500$ ms and $1$ s, the proposed scheme can still detect the splicing segments for carefully chosen threshold, instead of the trained ``Optimal Threshold" in Section \ref{sec:Audio_Source_Auth}. For instance, in Fig. \ref{fig:synthetic_data_splicing_det_diff_frame}(e), we can still obtain good detection and localization performance if the threshold is equal to $0.17$. Reducing the frame size further deteriorates both the performance of detection accuracy and localization granularity (see also Fig. \ref{fig:synthetic_data_splicing_det_diff_frame}(f)). It has been observed through extensive experimentation that for the data sets used here the frame size of $1$s may be a good compromise and the optimal frame size should be determined according to the application at hand.

\begin{figure}[htb]
  \centering
  \includegraphics[width=3.5in,height=2.8in]{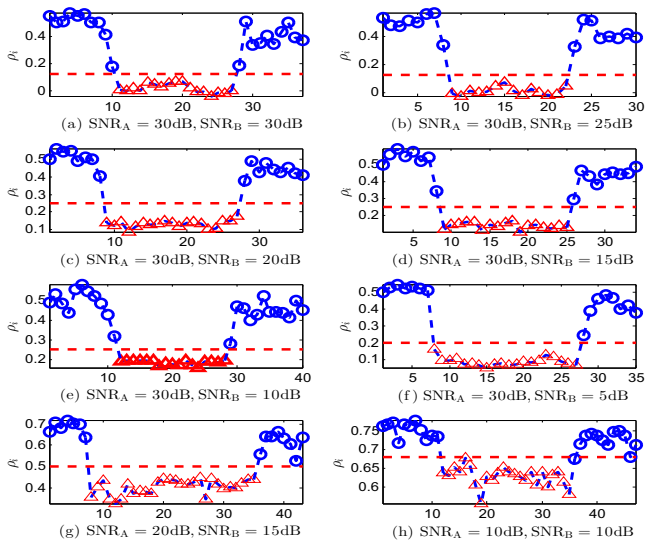}\\
  \caption{Detecting performance of the proposed scheme under with various SNRs. The frame size is 3s for all runs. Blue circles and red triangles represent the frames from $RS_A$ and $RS_B$, respectively. The red lines are the corresponding ``\emph{optimal}" decision thresholds. }
  \label{fig:synthetic_data_splicing_det_SNRs}
\end{figure}

The effect of noise on the proposed scheme has also been investigated. To this end, two randomly generated RIRs $H_A$ and $H_B$ were convolved with two randomly selected speeches from the test data set, resulting in reverberant speeches $RS_A$ and $RS_B$, respectively. Noisy speech samples were generated by adding white Gaussian noise with various Signal-to-Noise Ratios (SNRs) to $RS_A$ and $RS_B$, and noisy forged recordings were generated by inserting $RS_B$ in the middle of $RS_A$. A total of 1344  noisy spliced audio recordings were obtained by varying SNR values. Fig. \ref{fig:synthetic_data_splicing_det_SNRs} shows the detection results of our proposed scheme under various SNR values. For each forged sample, roughly, the $13^{th}$ through $44^{th}$ frames belong to $RS_B$, others come from $RS_A$. $SNR_A$ and $SNR_B$ represent the SNRs of reverberant speeches $RS_A$ and $RS_B$, respectively. It can be observed from Fig. \ref{fig:synthetic_data_splicing_det_SNRs}(a)-(f), if the SNR of either speech recording (say $RS_A$) is kept relatively high, then the proposed scheme can detect and localize the inserted frames with very high accuracy, regardless of the SNR of another speech ($RS_B$). As acoustic environment signature is generally modeled as a combination of the RIR and the background noise, the divergence of background noise is also a trace of counterfeit \cite{pan:2012, sohaib:2010}. It has been observed that by decreasing the SNR level, detection performance of the proposed scheme deteriorates. It can also be observed from  Fig. \ref{fig:synthetic_data_splicing_det_SNRs}(g)-(h) that the FPR performance of the proposed scheme increases gradually as SNR decreases. However, even under low SNR conditions (e.g., $10$dB), the proposed scheme exhibits good detection and localization performance.

\subsubsection{Results on Real World Data}

Effectiveness of the proposed audio splicing detection method has also been verified on real-world speech data. To achieve this goal, a data set consisting of 60 real world speech recordings of three speakers reading five different texts made in four acoustically different environments is used. In addition, the audio recordings made in the same environment with the same speaker (different text contents) is downsampled to 16 kHz and concatenated to generate each audio recording of 90s to 120s duration. For faircomparison, each forged audio is created by inserting audio segment (from other recording) into the middle of another recording.

\begin{figure}[htb]
  \centering
  \includegraphics[width=3.5in,height=3.2in]{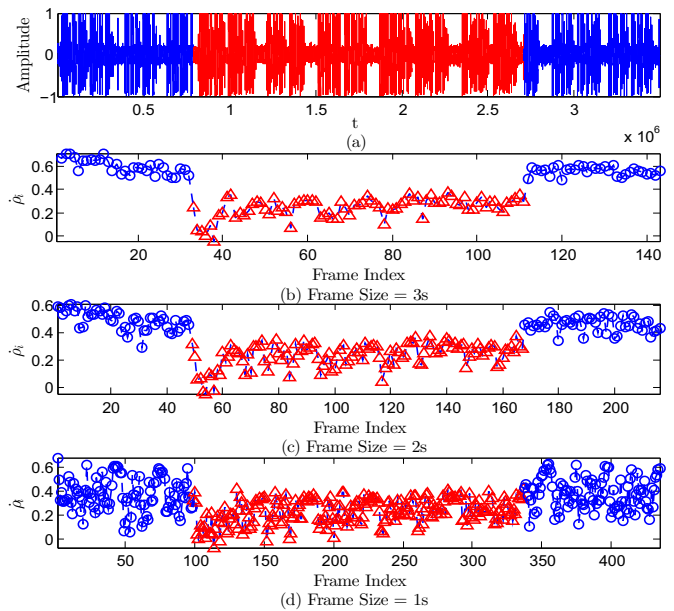}\\
  \caption{Detecting performance of the proposed scheme on real-world data with various frame sizes: (b)3s, (c)2s, and (d):1s. Blue circles and red triangles represent the frames from {E1S1} and {E3S2}, respectively }
  \label{fig:real_data_det_examp}
\end{figure}

Fig. \ref{fig:real_data_det_examp} shows the detection results of the proposed scheme on real world recordings created by inserting speech recording of the $1^{st}$ speaker in the $1^{st}$ environment (red part in Fig. \ref{fig:real_data_det_examp}(a)) in the middle of speech recording of the $2^{nd}$ speaker in the $3^{rd}$ environment [blue part in Fig. \ref{fig:real_data_det_examp}(a)]. It can be observed from Fig. \ref{fig:real_data_det_examp}(a) that all three segments contain strong but similar background noise and splicing did not introduce any visual artifacts. The resulting audio is analyzed using the proposed splicing  detection and localization method. For frame size $\geq 1s$, the optimal decision threshold was set to be $0.4$, i.e., $T=0.4$. It has been observed however that detection performance gradually deteriorates [e.g., both the FPR and FNR (the probability of query audio is labeled as \textit{authentic} when in fact it is \textit{forged}.) increases] for smaller frame sizes. This fact is illustrated in  Fig. \ref{fig:real_data_det_examp}(c)$\&$(d). It can be observed that the proposed scheme is still capable of detecting/localizing the inserted frames for frame sizes 2s and 1s. However, optimal decision threshold selection becomes very difficult for smaller framework size. Pattern recognition methods such as support vector machine\cite{SVM} may be applied to address these limitations. Nevertheless, discussion on this is not the focus of this paper.

It has also been observed through experimentation that some of the recordings in the data set contain stronger background noise than the signal shown in Fig. \ref{fig:real_data_det_examp}(a), which deteriorates the accuracy of environment signature estimation, and hence degrades the performance of our proposed scheme. The noise effect can be reduced by using long-term average but at the cost of localization resolution. Experimental results indicate that frame sizes between 2s and 3s resulted in good detection performance for median noise level, and for strong background noise, frame sizes over 5s may be needed for acceptable performance. Similar results were obtained for other forged recordings.

%Fig. \ref{fig:real_data_det_examp_multiple} shows the detection results of the proposed scheme on a more general case, where the audio clip is spliced by assembling 5 segments. The frames marked by blue circles are captured in one environment but with different speech contents. And the frames marked by red triangles are captured in other different environments. It can be observed that the proposed scheme can detect and localize the frames which are captured in environments different from that of the first segment. Similar results can be obtained from other tests.
%\begin{figure}[htb]
%  \centering
%  % Requires \usepackage{graphicx}
%  \includegraphics[width=3.5in,height=3in]{real_world_data_det_example_multiple_splice_v2.eps}\\
%  \caption{Detecting performance of the proposed scheme on real-world data with various frame sizes: (a)2s, (b)3s, and (c):4s. The query audio was spliced by assembling 5 segments.}
%  \label{fig:real_data_det_examp_multiple}
%\end{figure}

\begin{figure}[htb]
  \centering
  \includegraphics[width=2.8in,height=2in]{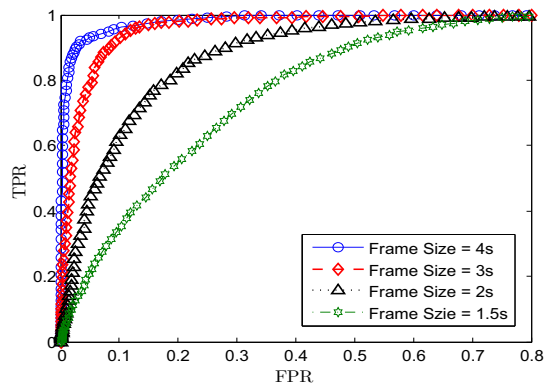}\\
  \caption{ROC curves of the proposed scheme on real-world data with various frame sizes.}
  \label{fig:real_data_det_ROC}
\end{figure}

In our next experiment, the detection accuracy of the proposed scheme for the real-world data set is evaluated. For this experiment, spliced audio is generated from two randomly selected audio recordings of different speakers made in different environments from the real-world data set. Two quantitative measures, TPR and FPR, are used to evaluate the frame-level detection performance. To achieve a smooth ROC plot, the TPR/FPR pair is computed by averaging over 50 runs of different frame sizes. Fig. \ref{fig:real_data_det_ROC} shows the resulting ROC curves computed at different frame sizes. It can be observed from  Fig. \ref{fig:real_data_det_ROC} that the overall detection accuracy improves for larger frame sizes. This is expected as larger frame sizes result in more accurate and stable estimation of environment signature. In addition, it can also be observed from Fig. \ref{fig:sythetic_data_ROC} and Fig. \ref{fig:real_data_det_ROC} that, the detection performance of the proposed scheme deteriorates in the presence of ambient noise. Detection performance in the presence of ambient noise can be improved by using larger frame sizes but at the cost of lower forgery localization accuracy.

\subsubsection{Robustness to MP3 Compression}

In this experiment, we tested the robustness of our scheme against MP3 compression attacks. To this end, audio recordings were compressed into MP3 format using FFMPEG \cite{ffmpeg} with bitrates $R\in\{16, 32, 64, 96, 128, 192, 256\}$ kbps. Spliced audio recordings were generated by randomly selecting compressed audio recordings captured in different environments. It is important to mention that each audio segment used to generate forged audio was compressed to MP3 format with the same set of parameters, e.g., the same bit-rate, etc.
%Otherwise, the spliced audio can be easily detected according to the trace of inconsistent compression history\cite{Fan:2003} or double compression\cite{Liu:2010}\cite{Qiao:2010}, which is not our focus in this paper.

For forgery detection, MP3 compressed forged audio is decoded to the wav format, which is then analyzed using the proposed method. Each experiment was repeated 50 times and detection performance was averaged over 50 runs to reduce the effect of noise. Table \ref{tab:real_world_data_mp3_att_det} shows the detection accuracy (the probability of correctly identifying the inserted and original frames) of the proposed scheme under MP3 attacks for various bit-rates and frame sizes.

\begin{table}[thbp]
  \centering
  \caption{Detection accuracies of the proposed scheme under MP3 attacks with various bit-rates and different frame sizes}
  \begin{tabular}{|c|c|c|c|c|c|c|c|}
    \hline
    Frame & \multicolumn{7}{c|}{MP3 Compression Bit Rate (\textit{bps})} \\ \cline{2-8}
    Size & $16k$ & $32k$ & $64k$ & $96k$ & $128k$ & $192k$ & $256k$ \\ \hline
    $4s$ & 0.86 & 0.92 & 0.93 & 0.93 & 0.93 & 0.93 & 0.94 \\ \hline
    $3s$ & 0.89 & 0.92 & 0.92 & 0.93 & 0.93 & 0.93 & 0.94 \\ \hline
    $2s$ & 0.90 & 0.92 & 0.91 & 0.91 & 0.93 & 0.93 & 0.93 \\ \hline
    $1s$ & 0.89 & 0.83 & 0.83 & 0.85 & 0.83 & 0.83 & 0.84 \\
    \hline
  \end{tabular}
  \label{tab:real_world_data_mp3_att_det}
\end{table}

The following observations can be made from Table \ref{tab:real_world_data_mp3_att_det}.

Firstly, the detection performance improves for higher bit-rates, which introduce less distortion. The detection accuracy keeps relatively unchanged (less than 1\% loss) for bit-rates $R \geq 32$ kbps. This indicates that the proposed scheme is robust to MP3 compression attack with bit-rate $R > 16$ kbps.

Secondly, the detection performance deteriorates (on average $7.6\%$) by changing the frame size from 2s to 1s. It has been verified in our previous experiments that smaller frame sizes result in relatively less accurate estimate of environment signature.

Finally, it can be observed that frame sizes less than 3s result in relatively unstable detection performance, which can be attributed to the random noise. In addition, for MP3 compression attack, the detection performance of the proposed scheme is relatively more sensitive to frame size. As the performance of forgery localization method depends on frame size used for analysis, forensic analyst can select an appropriate frame size for the application and test audio at hand. It has been observed that for the data set used, frame size $> 2s$ results in good compromise between the detection and localization accuracies.

\subsubsection{Comparison Results with Previous Works}

In our last set of experiments, performance of the proposed framework is compared to the existing splicing detection scheme \cite{pan:2012}, which uses inconsistency in local noise levels estimated from the query audio for splicing detection.

\begin{figure}[htb]
  \centering
  \includegraphics[width=3.5in,height=2.75in]{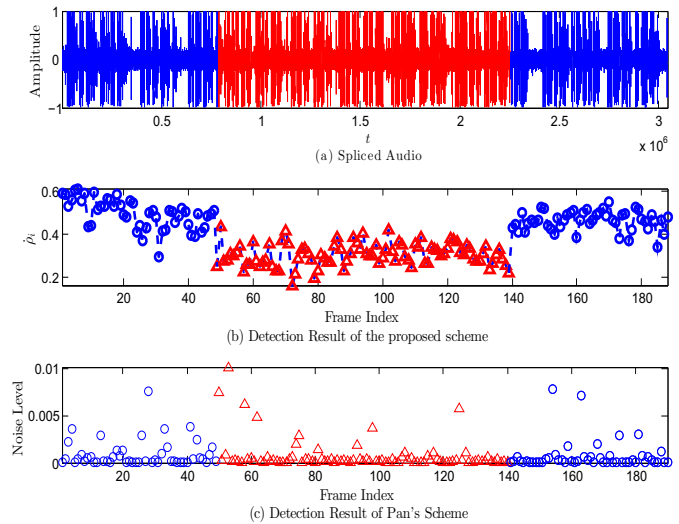}\\
  \caption{Detection performance of the proposed scheme and Pan's scheme \cite{pan:2012}. For fair comparison, frame size is set to 2s for both experiments. Red triangles represent the frames of recording of the $3^{rd}$ speaker in the $3^{rd}$ environment (E3S3) and blue circles represent frames of recording of the $1^{st}$ speaker in the $1^{st}$ environment (E1S1).}
  \label{fig:real_data_com_sim_noise_level}
\end{figure}

To this end, forged (or spliced) audio recording is generated by splicing audio recordings made in two different environments. More specifically, speech recording of the $3^{rd}$ speaker made in the $3^{rd}$ environment (E3S3) is inserted in the middle of speech recording of the $1^{st}$ speaker made in the $1^{st}$ environment (E1S1). Fig. \ref{fig:real_data_com_sim_noise_level}(a) presents the time-domain plot of the spliced audio assembled from E1S1 (blue) and E3S3 (red). It can be observed from Fig.\ref{fig:real_data_com_sim_noise_level}(a) that both original recordings, e.g., E1S1 and E3S3,  contain the same background noise level.  Fig. \ref{fig:real_data_com_sim_noise_level}(b) and Fig. \ref{fig:real_data_com_sim_noise_level}(c) show the frame-level detection performance of the proposed scheme and Pan's scheme \cite{pan:2012}, respectively. It can be observed from Fig. \ref{fig:real_data_com_sim_noise_level}(b)\&(c) that the proposed scheme is capable of not only detecting the presence of the inserted frames but also localizing these frames. On the other hand, Pan's scheme is unable to detect or localize the inserted frames. Inferior performance of Pan's scheme can be attributed to the fact that it only depends on the local noise level for forgery detection, which is almost the same in the forged audio used for analysis.

\begin{figure}[htb]
  \centering
  \includegraphics[width=3.5in,height=2.7in]{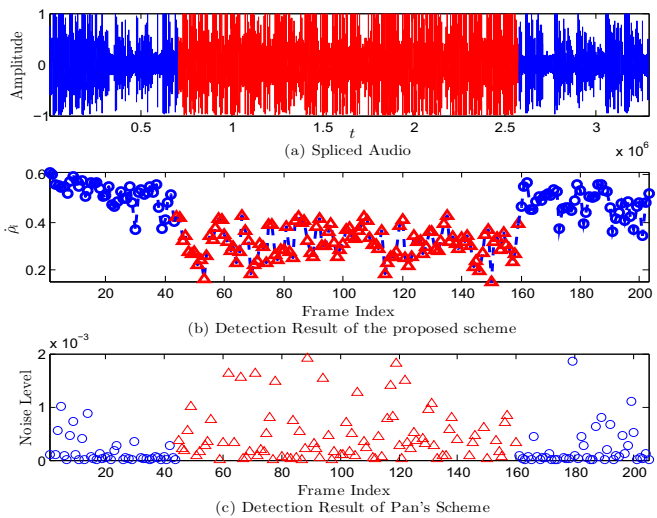}\\
  \caption{Detection performance of the proposed scheme and Pan's scheme \cite{pan:2012}. For fair comparison, the frame size is set to 2s for both experiments. Red triangles represent the frames of recording of the $3^{rd}$ speaker in the $2^{nd}$ environment (E2S3) and blue circles represent the frames of recording of the $2^{nd}$ speaker in the $1^{st}$ environment (E1S1).}
  \label{fig:real_data_com_diff_noise_level}
\end{figure}

To further demonstrate the superiority of our proposed scheme, a forged audio is generated by splicing together two speech recordings with different levels of background noise. More specifically, the speech recording of the $3^{rd}$ speaker made in the $2^{nd}$ environment (E2S3) is inserted at the middle of speech recording of the $2^{nd}$ speaker made in the $1^{st}$ environment (E1S2). Fig. \ref{fig:real_data_com_diff_noise_level}(a) shows the time-domain plot of the forged audio assembled from E1S2 (blue) and E2S3 (red). It can be observed that the segment from E2S3 exhibits higher noise level than the segments from E1S2. The resulting forged recording is analyzed using both the proposed scheme and Pan's scheme \cite{pan:2012}. Fig. \ref{fig:real_data_com_diff_noise_level}(b) and Fig. \ref{fig:real_data_com_diff_noise_level}(c) illustrate the the plots of frame-level detection performance of these two schemes, respectively. It can be observed from Fig. \ref{fig:real_data_com_diff_noise_level}(b) that the proposed scheme has successfully detected and localized the inserted frames with much higher accuracy than the result in \cite{pan:2012}. In addition, it can be observed from Fig. \ref{fig:real_data_com_diff_noise_level}(c) that Pan's scheme is unable to localize the inserted frames. Further more, Pan's scheme achieves a detection accuracy close to $0.6$, which is far lower than the accuracy achieved by the proposed scheme, $0.93$. Finally, the detection performance of both schemes are also evaluated for larger frame sizes. As expected performance of the proposed improved significantly for frame size $> 2s$. However, no significant improvement is observed for Pan's scheme.

In our final experiment, the performance of the proposed scheme and Pan's scheme \cite{pan:2012} is evaluated on forged audio clips. Forged audio clips were generated by randomly selecting audio clips (of approximately 3s duration) from TIMIT  and modifying each of them by adding white Gaussian noise ($SNR\in\{15\text{dB}, 20\text{dB}, 25\text{dB}, 30\text{dB}\}$) into the first 50\% of the frames (Other frames are kept to be clean.). It can be observed that Pan's scheme performs reasonably well for this forged data set. However, when this experiment is repeated for audio clips of 30 second duration, FPRs, for Pan's scheme increase significantly. This experiment implies that generalization of Pan's scheme \cite{pan:2012} requires further verification.

Performance comparison of the proposed scheme to Pan's scheme \cite{pan:2012} can be summarized as follows: (i) superior performance of the proposed scheme can be attributed to RIR which is an acoustic environment related signature and is more reliable than the ambient noise level alone; (ii) the proposed scheme models acoustic environment signature by considering both the background noise (including the level, type, spectrum, etc.) and the RIR (see (\ref{eq:est_RIR})); and (iii) the proposed scheme relies on the acoustic channel impulse response, which is relatively hard to manipulate as compare to the background noise based acoustic signature.

\section{Conclusion}\label{sec:diss}
In this paper, a novel method for audio splicing detection and localization has been proposed. The magnitude of acoustic channel impulse response and ambient noise have been considered to model the intrinsic acoustic environment signature. The acoustic environment signature is jointly estimated using the spectrum classification technique. Distribution of the correlation coefficient between the query audio and the reference audio has been used to determine the optimal decision threshold, which is used for the frame-level forgery detection. A frame with similarity score less than the optimal threshold is labeled as spliced and as authentic otherwise. For splicing localization, a local correlation based on the relationship between adjacent frames has been adopted to reduce frame-level detection errors. The performance of the proposed scheme has been evaluated on two data sets with numerous experimental settings. Experimental results has validated the effectiveness of the proposed scheme for both forgery detection and localization. Robustness of the proposed scheme to both MP3 compression and additive noise attacks has also been tested. Performance of the proposed scheme has also been compared against the state-of-the-art for audio splicing detection \cite{pan:2012}. Performance comparison has indicated that the proposed scheme outperforms the approach presented in \cite{pan:2012}. The proposed scheme may be a good candidate for the practical digital audio forensics applications.

\section*{Acknowledgment}
This work is supported by 2013 Guangdong Natural Science Funds for Distinguished Young Scholar (S2013050014223).

\ifCLASSOPTIONcaptionsoff
  \newpage
\fi
\bibliographystyle{IEEEtran}
\bibliography{RefAudioForensics}

\vfill


\end{document}